\documentclass[lettersize,journal]{IEEEtran}
\usepackage{amsmath,amsfonts}
\usepackage{array}
\usepackage[caption=false,font=footnotesize,labelfont=sf,textfont=sf]{subfig}
\usepackage{textcomp}
\usepackage{stfloats}
\usepackage{url}
\usepackage{verbatim}
\usepackage{graphicx}
\def\BibTeX{{\rm B\kern-.05em{\sc i\kern-.025em b}\kern-.08em
    T\kern-.1667em\lower.7ex\hbox{E}\kern-.125emX}}
\usepackage{balance}
\usepackage{cite}
\usepackage{CJKutf8}
\usepackage{xcolor}
\usepackage{enumitem}
\usepackage{amsmath}
\usepackage{algpseudocode}
\usepackage{graphicx}
\usepackage{amssymb}
\usepackage{caption}
\begin{document}

\title{An LLM-based Self-Evolving Security Framework for 6G Space-Air-Ground Integrated Networks}

\author{Qi Qin, Xinye Cao, Guoshun Nan, \textit{Member, IEEE,} Sihan Chen, Rushan Li, Li Su, Haitao Du, Qimei Cui, \textit{Senior Member, IEEE,} Pengxuan Mao, Xiaofeng Tao, \textit{Senior Member, IEEE,} Tony Q.S. Quek, \textit{Fellow, IEEE}
\thanks{Qi Qin, Xinye Cao, Guoshun Nan, Sihan Chen, Rushan Li, Qimei Cui, and Xiaofeng Tao are with the National Engineering Research Center for Mobile Network Technologies, Beijing University of Posts and Telecommunications, China, also with Beiyou Shenzhen Institute. (Corresponding author: Guoshun Nan.)}
\thanks{Li Su and Haitao Du are professorate senior engineers at China Mobile Research Institute, China.}
\thanks{Pengxuan Mao is with Terminus Technologies Co., Ltd.}
\thanks{T. Q. S. Quek is with the Singapore University of Technology and Design, Singapore 487372, and also with the Yonsei Frontier Lab, Yonsei University, South Korea.}
\thanks{Qi Qin and Xinye Cao are equally contributed.}}

\maketitle

\begin{abstract}
Recently emerged 6G space-air-ground integrated networks (SAGINs), which integrate satellites, aerial networks, and terrestrial communications, offer ubiquitous coverage for various mobile applications. 
However, the highly dynamic, open, and heterogeneous nature of SAGINs poses severe security issues. Forming a defense line of SAGINs suffers from two preliminary challenges: 1) accurately understanding massive unstructured multi-dimensional threat information to generate defense strategies against various malicious attacks, 2) rapidly adapting to potential unknown threats to yield more effective security strategies.
To tackle the above two challenges, we propose a novel security framework for SAGINs based on Large Language Models (LLMs), which consists of two key ingredients LLM-6GNG and 6G-INST. 
Our proposed LLM-6GNG leverages refined chain-of-thought (CoT) reasoning and dynamic multi-agent mechanisms to analyze massive unstructured multi-dimensional threat data and generate comprehensive security strategies, thus addressing the first challenge.
Our proposed 6G-INST relies on a novel self-evolving method to automatically update LLM-6GNG, enabling it to accommodate unknown threats under dynamic communication environments, thereby addressing the second challenge. 
Additionally, we prototype the proposed framework with ns-3, OpenAirInterface (OAI), and software-defined radio (SDR). Experiments on three benchmarks demonstrate the effectiveness of our framework. The results show that our framework produces highly accurate security strategies that remain robust against a variety of unknown attacks. 
We will release our code to contribute to the community.

\end{abstract}


\section{Introduction}

\begin{figure}[!htbp]
\vspace{-4mm}
  \centering
  \includegraphics[width=0.5\textwidth]{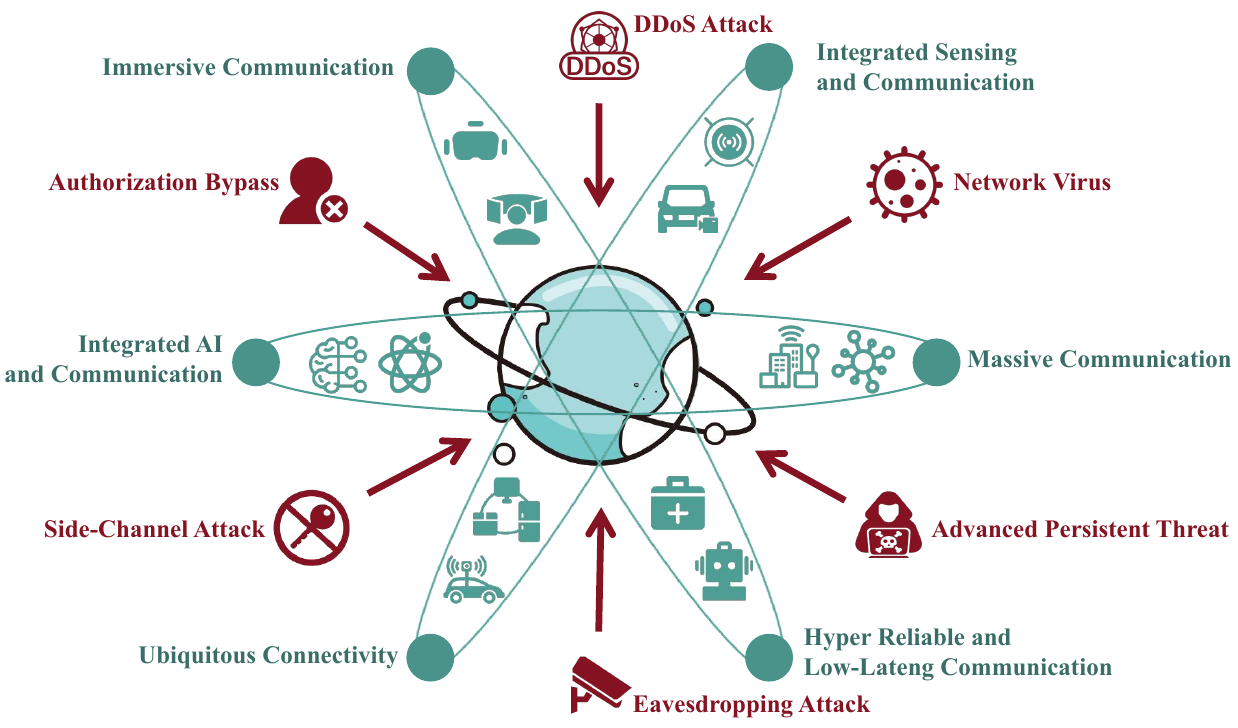}
  \caption{Illustration of 6G usage scenarios and the corresponding threats. 6G introduces new usage scenarios such as artificial intelligence and communication, ubiquitous connectivity, and massive communication, which bring more security threats including DDoS and advanced persistent threat (APT).}
  \label{fig:fig1}
  \vspace{-7mm}
\end{figure}

\IEEEPARstart{A}{s} the International Telecommunication Union (ITU) articulates the vision for 6G networks, the characteristics and usage scenarios of 6G are becoming increasingly clear \cite{c23}. The core usage scenarios for 6G SAGINs include artificial intelligence and communication, ubiquitous connectivity, and massive communication. Meanwhile, the 6G communication architecture is inherently cross-layer and cross-domain \cite{c22, c14}. Cross-layer implies that various layers within the communication system, such as the Air Interface Layer, Access Layer, Core Network Layer, and Application Layer, will transcend conventional boundaries to facilitate information exchange and resource sharing. Cross-domain involves global coverage, seamless communication, and emergency recovery across satellite systems, aerial networks, and terrestrial networks. 
Furthermore, the swift advancement of 6G technology has introduced a multitude of emerging applications, including autonomous driving, telemedicine, and smart cities. These innovations have markedly improved both industrial productivity and the quality of life for citizens.

While 6G SAGINs introduce novel usage scenarios, the cross-layer and cross-domain features, along with emerging applications of 6G, expand attack surfaces and increase complexity in detecting and mitigating security threats, as illustrated in Figure \ref{fig:fig1}. For instance, malicious actors can launch DDoS attacks to disrupt services such as autonomous vehicles and smart urban infrastructures. Additionally, unauthorized individuals might eavesdrop on wireless signals, thereby leaking sensitive data transmitted in the space-to-ground link. Therefore, 6G security frameworks are needed to ensure the security of 6G SAGINs.

Towards 6G security frameworks, numerous studies have been conducted to address emerging challenges and ensure effective protection in 6G networks \cite{c16,c17,c18}. Wang et al. proposed the SIX-Trust framework, a multi-layer trust model for 6G, focusing on Sustainable Trust, Infrastructure Trust, and Xenogenesis Trust to address security challenges \cite{c19}. Wen et al. introduced the 6G-XSec framework, which combines unsupervised deep learning-based anomaly detection with LLMs to monitor, analyze, and explain security threats at the edge of OpenRAN-based 6G networks \cite{c20}. Mekrache et al. introduced a novel approach for Zero-Touch Network and Service Management (ZSM) in 6G, combining AI for anomaly detection, explainable AI (XAI) for identifying root causes, and LLMs for providing user-friendly explanations and corrective actions \cite{c21}.
However, previous works have not investigated methods for handling massive unstructured multi-dimensional threat information and generating comprehensive security strategies in the context of 6G SAGINs. Additionally, there has been limited research on how to enable security frameworks to continuously self-evolve in order to address emerging and unknown threats.

To fulfill the aforementioned gap, we propose a 6G security framework to enable massive unstructured multi-dimensional threat information processing and comprehensive security strategy generation in 6G SAGINs. We also implement a self-evolving method to empower LLMs to defend against unknown threats. Additionally, we build a semi-physical 6G Simulator based on ns-3, OAI, and SDR to validate the effectiveness of the security architecture. The contributions of this paper are summarized as follows:

\begin{itemize}[leftmargin=*]

    \item We present the LLM-based 6G network guard (LLM-6GNG), a novel method that applies multi-agent LLMs with CoT reasoning to enhance 6G SAGINs security. Specifically, massive multi-dimensional threat information can be effectively condensed, classified, and extracted through our LLM-6GNG. Furthermore, comprehensive security strategies are generated in our LLM-6GNG to address complex and dynamic security threats under the 6G communications.

    \item To address emerging threats in 6G networks, we propose the 6G-Instruction (6G-INST) method, which automatically updates training datasets, thereby enabling the self-evolution of our LLM-6GNG. To the best of our knowledge, we are the first to explore the self-evolving security strategy generation method.

    \item A 6G Simulator based on OAI and ns-3 is developed to generate a 6G SAGINs threat information dataset. Experiments on three datasets related to 6G network threats validate the performance of our LLM-6GNG integrated with the 6G-INST against unknown threats and demonstrate the potential of our framework effectively addressing diverse and dynamic security challenges in 6G SAGINs. Additionally, we visualize the work procedure in the case study.

\end{itemize}

\section{System Overview}
\begin{figure*}[!htbp]
\vspace{-8mm}
  \centering
  \includegraphics[width=\textwidth]{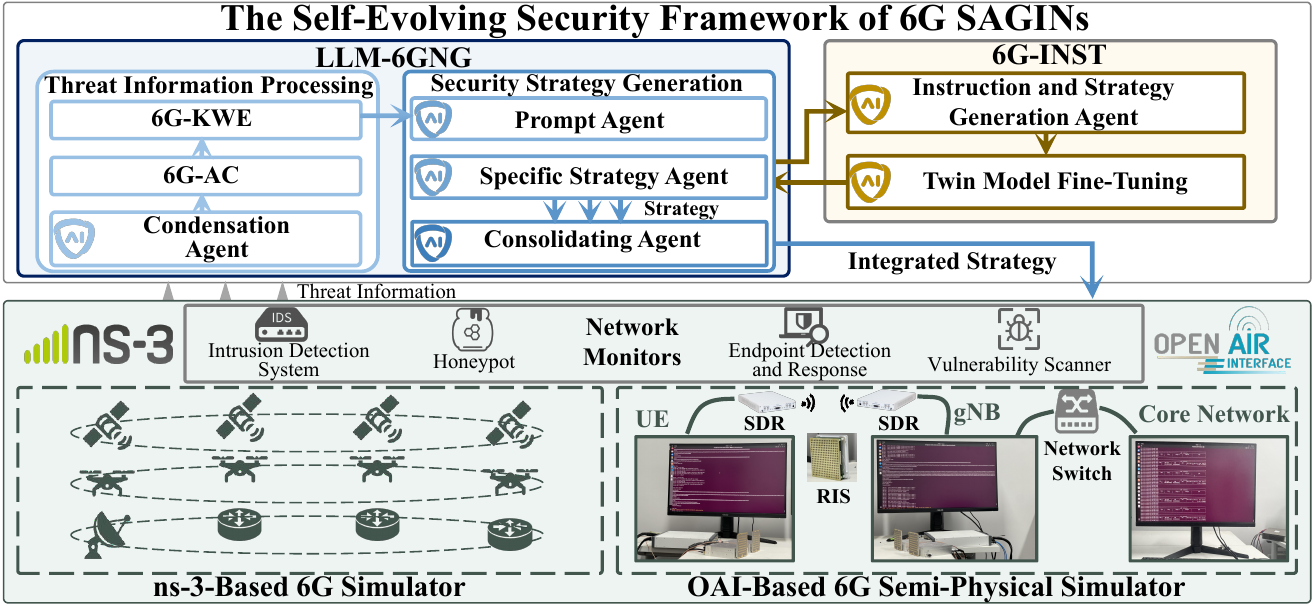}
  \caption{Illustration of our proposed self-evolving security framework of 6G SAGINs and the 6G Simulator. The proposed LLM-6GNG receives threat information from our 6G Simulator, and feeds back security strategies to the 6G Simulator. Meanwhile, our 6G-INST collects data from the proposed LLM-6GNG, trains the Twin Specific Strategy Agent, and updates the LLMs.}
  \label{fig:fig2}
  \vspace{-7mm}
\end{figure*}

To simulate a more realistic 6G SAGINs communication system and address emerging threats, we construct the 6G Simulator and the self-evolving security framework of 6G SAGINs as depicted in Figure \ref{fig:fig2}. Our system consists of three components: 6G Simulator, LLM-6GNG, and 6G-INST. Threat information collected in the 6G Simulator is passed to the LLM-6GNG, which analyzes the threat information and generates corresponding security strategies in real-time. Concurrently, the 6G-INST empowers our LLM-6GNG with self-evolution capability.

\subsection{Our proposed 6G Simulator}

Our 6G Simulator consists of three main components: an ns-3-based simulator, an OAI-based semi-physical simulator, and a monitoring system.

\subsubsection{\textbf{ns-3-Based 6G Simulator}}
We develop a 6G space-air-ground network simulator based on ns-3 including three components: the satellite system, the aerial network, and the terrestrial network.
For the space-based network, we construct the satellite system including GEO and LEO satellites, and develop a physical layer protocol based on DVB-S2X. For the air-based network, we deploy UAVs equipped with WLAN protocols and configure the random walk mobility pattern to enhance network dynamics and coverage. For the ground-based network, the module comprises ground base stations and user equipment, and implements the LTE protocol stack for communications.

\subsubsection{\textbf{OAI-Based 6G Semi-physical Simulator}}
Our OAI-based 6G Simulator is a highly integrated communication platform, which allows for comprehensive tests of security performance within the 6G system. 
The simulator encompasses the air interface layer, access layer, and core network layer, enabling detailed simulation and analysis of attacks and detection mechanisms at each layer of the communication system.
We deploy the customizable OAI UE and base station software equipped with software defined radio (SDR) to facilitate the transmission and reception of signals. The system is also equipped with upconverters and downconverters, which are crucial for efficiently converting the original signals to and from the required millimeter-wave frequencies, respectively. Additionally, the system innovatively incorporates a reconfigurable intelligent surface (RIS). 
Finally, the core network module is simulated using the OAI core network software. We also implement custom modifications to the network functions of OAI, adding a significant number of security functions to enhance the 6G system.

\subsubsection{\textbf{Cyber Attack Simulator and Network Monitor}}
We create a Cyber Attack Simulator and deploy it across multiple nodes in our 6G Simulator to simulate common attack traffic such as distributed denial of service (DDoS) attacks, DoS attacks, web attacks, as well as penetration attacks, brute force attacks, and deception. 
At the same time, to monitor the current state of the 6G SAGINs in real-time, we deploy anomaly detection modules at various layers and domains of our 6G Simulator, including intrusion detection systems (IDS), intelligent gateways, vulnerability scanning, air interface anomaly detection, endpoint detection and response (EDR), and honeynets. So that we can detect the multi-dimensional attack information generated by the Cyber Attack Simulator in real-time.

\subsection{Our proposed LLM-6GNG and 6G-INST}
Our LLM-6GNG consists of two modules: the threat information processing module, which extracts key information from large amounts of unstructured threat information, and the security strategy generation module, which generates security strategies based on the extracted information. Our 6G-INST is used to enable the self-evolution of LLM-6GNG by collecting data, generating new threat data and strategies, and fine-tuning the system's parameters. Detailed descriptions of LLM-6GNG and 6G-INST will be provided in Sections \ref{sec:LLM-6GNG} and \ref{sec:self-Evolving}.

\section{Our Proposed LLM-6GNG}
\label{sec:LLM-6GNG}
Our LLM-6GNG comprises two modules, namely the threat information processing module and the security strategy generation module. As shown in Figure \ref{fig:fig3}, we illustrate the procedure of our LLM-6GNG. The procedure of the threat information processing module is depicted in the upper half of the middle layer of Figure \ref{fig:fig3}. The threat information collected from the 6G Simulator is processed by the Condensation Agent, the 6G attack correlation (6G-AC) algorithm, and the 6G key information extraction (6G-KWE) algorithm to extract keywords from the threat information. 
The procedure of the security strategy generation module is depicted in the lower half of the middle layer of Figure \ref{fig:fig3}. After being processed by the threat information processing module, we utilize the Prompt Agent, the Specific Strategy Agent, and the Consolidating Agent to analyze the processed key data and generate security strategies, which are then fed back to our 6G Simulator. 
In our LLM-6GNG, CoT reasoning and the multi-agent collaboration mechanism are employed, which effectively analyze threat information and generate accurate security strategies.
To meet the real-time requirements for security strategy generation, we chose Llama3-8b to build LLM-6GNG, ensuring its high operational efficiency. Next, we will provide a detailed introduction to LLM-6GNG.

\subsection{\textbf{The Condensation Agent}}
In 6G SAGINs, threat information may be collected from a variety of different sources. These threat data are not standardized, and we cannot use the same methods to extract their key information. Therefore, we need the Condensation Agent to perform initial processing of the threat information. The Condensation Agent is capable of transforming threat information collected from multiple sources into specific threat descriptions and main features. This agent provides clear and organized key information, which facilitates the subsequent components in using regular expressions to extract the key features of the threat information as well as other processing.

\subsection{\textbf{6G Attack Correlation}}

The 6G attack correlation algorithm processes the aggregated data within each subnet for subsequent processing. It regroups the threat information based on the semantic and feature similarity between them, ultimately generating correlated information. As shown in Figure \ref{fig:fig3}, for intra-network communications, the procedure of 6G-AC is as follows:

\noindent \textbf{1. Semantic similarity and feature similarity:} 
Firstly, for each pair of threat information, we use a bidirectional encoder representation from transformers (BERT) model to analyze the semantic similarity between their threat descriptions. Then, we use regular expressions to extract key features from all these pairs of threat information and calculate the similarity between their key features.

\noindent \textbf{2. Similarity integration and classification:} Next, we combine semantic similarity and feature similarity to calculate a comprehensive similarity score through a weighted sum. If this score exceeds a set threshold, the related threat information will be merged together. Ultimately, we obtain a collection of threat information groups.

For inter-network communications, as shown in Figure \ref{fig:fig3}, we aggregate threat information from all subnets involved in the inter-network communications. The aggregated information is then processed using the 6G-AC algorithm to achieve effective attack clustering across networks.

\subsection{\textbf{6G Key Information Extraction}}

The 6G-KWE algorithm assigns weights based on the presence of key terms within the threat information and Term-Frequency Inverse Document Frequency (TF-IDF) matrix, thereby filtering out the most important keywords from a multitude of threat information. 

In processing the threat information, we first extract key information from each piece of threat data in the group based on a predefined rule pattern. Key information is then converted into a TF-IDF matrix, where each value represents the importance of the key information within the specific threat data. Next, we multiply the TF-IDF matrix by the weight matrix to obtain the final weighted matrix. Finally, we analyze the final weighted matrix, while also identifying the most frequent content within the threat information to extract the most important key information.

\subsection{\textbf{Security Strategy Generation}}

After being processed by the threat information processing module, we utilize the advanced chain-of-thought \cite{c12} reasoning and summarization capabilities of LLM to analyze the processed key data and generate security strategies.

\noindent \textbf{1. Initialization and data preparation:} Firstly, we use the Prompt Agent to generate specific prompts and threat descriptions for each group of key threat data. Each group of prompts and corresponding threat descriptions is then transmitted to the Specific Strategy Agent to generate security strategies.

\noindent \textbf{2. Strategy generation:} Subsequently, the Specific Strategy Agent analyzes the received threat descriptions based on the generated prompts and conducts an in-depth evaluation to determine how to address the threats. Finally, it will generate security strategies against the specific threats.

\noindent \textbf{3. Strategy consolidation:} At last, the Consolidating Agent aggregates and integrates the diverse strategies produced by Specific Strategy Agents. Through comprehensive analysis of these strategies, it constructs a well-rounded and overarching integrated strategy, ensuring the coordination and consistency of the overall security strategy.

\begin{figure*}[!htbp]
\vspace{-8mm}
  \centering
  \includegraphics[width=\textwidth]{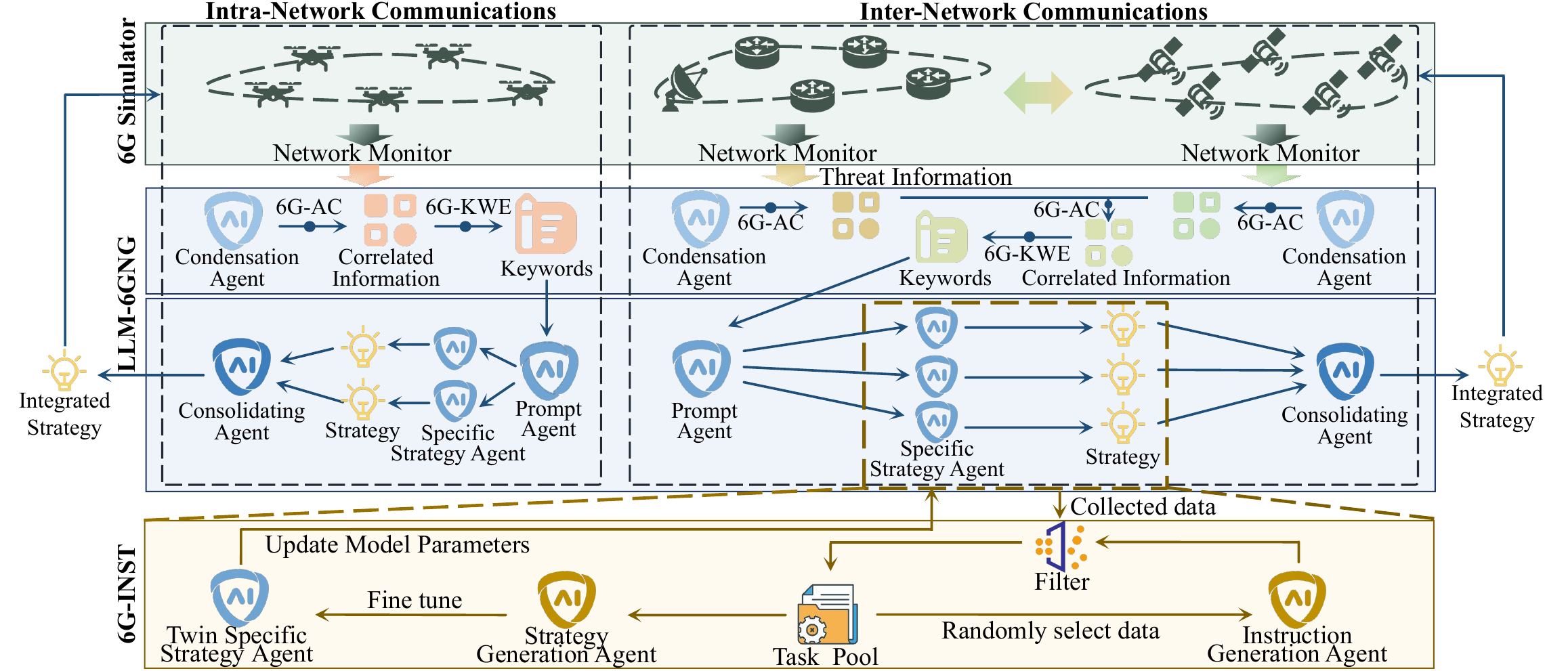}
  \caption{Illustration of our LLM-6GNG and 6G-INST. The top layer provides an example of threat information from the three subnets within our 6G Simulator. The middle part of the diagram describes the process of the LLM-6GNG, with the left side depicting the scenario of intra-network communication, and the right side showing the scenario of inter-Network Communication. The bottom layer of the diagram describes the process by which our 6G-INST automatically generates new datasets to assist the LLM-6GNG in its self-evolution.}
  \label{fig:fig3}
  \vspace{-6mm}
\end{figure*}

\section{Our Proposed 6G-INST}
\label{sec:self-Evolving}

As shown in the bottom layer of Figure \ref{fig:fig3}, we illustrate the procedure of our proposed 6G-INST. Our 6G-INST can collect and filter data from the Specific Strategy Agent in real-time and can automate the expansion of training datasets through the Instruction Generation Agent and Strategy Generation Agent. Through our proposed 6G-INST, we can address the high cost of collecting instruction datasets and use the expanded training datasets to fine-tune our LLM-6GNG, enabling its self-evolution.

\subsection{Prompt and Threat Generation}
In the process of expanding the instruction training dataset, the first step is to generate more prompts and threat information. Next, we will explain this process in detail.

\noindent \textbf{1. Preparation of the seed task set:}
First of all, we prepare 100 manually-written instructions, each comprising prompts, threat information, and strategies. These 100 tasks are utilized to initialize the task pool.

\noindent \textbf{2. Real-time data collection:}
Meanwhile, prompts and threat information encountered by the Specific Strategy Agent are added to the task pool after being filtered. They will be used to generate more instructions in subsequent processes.

\noindent \textbf{3. Prompt and threat information generation:}
Next, we randomly select 8 instructions from the task pool to generate new prompts and threat information by the Instruction Generation Agent. In the first round, all 8 instructions are manually written, and in subsequent rounds, we mix 5 manually written and 3 newly generated instructions for diversity. If real-time data is available, we select four manually written instructions, two newly generated, and two from the Specific Strategy Agent to generate new training data.

\noindent \textbf{4. Filtering and post-processing:}
To foster diversity, we only add collected instructions and generated instructions to the task pool if their ROUGE-L similarity score with all existing instructions is below 0.7. We also exclude instructions containing specific keywords, as these are typically not processable by LLMs.

After a full round of generation is completed, the process loops back to the third step, continuously generating new training data.

\begin{figure*}[!htbp]
\vspace{-8mm}
  \centering
  \includegraphics[width=\textwidth]{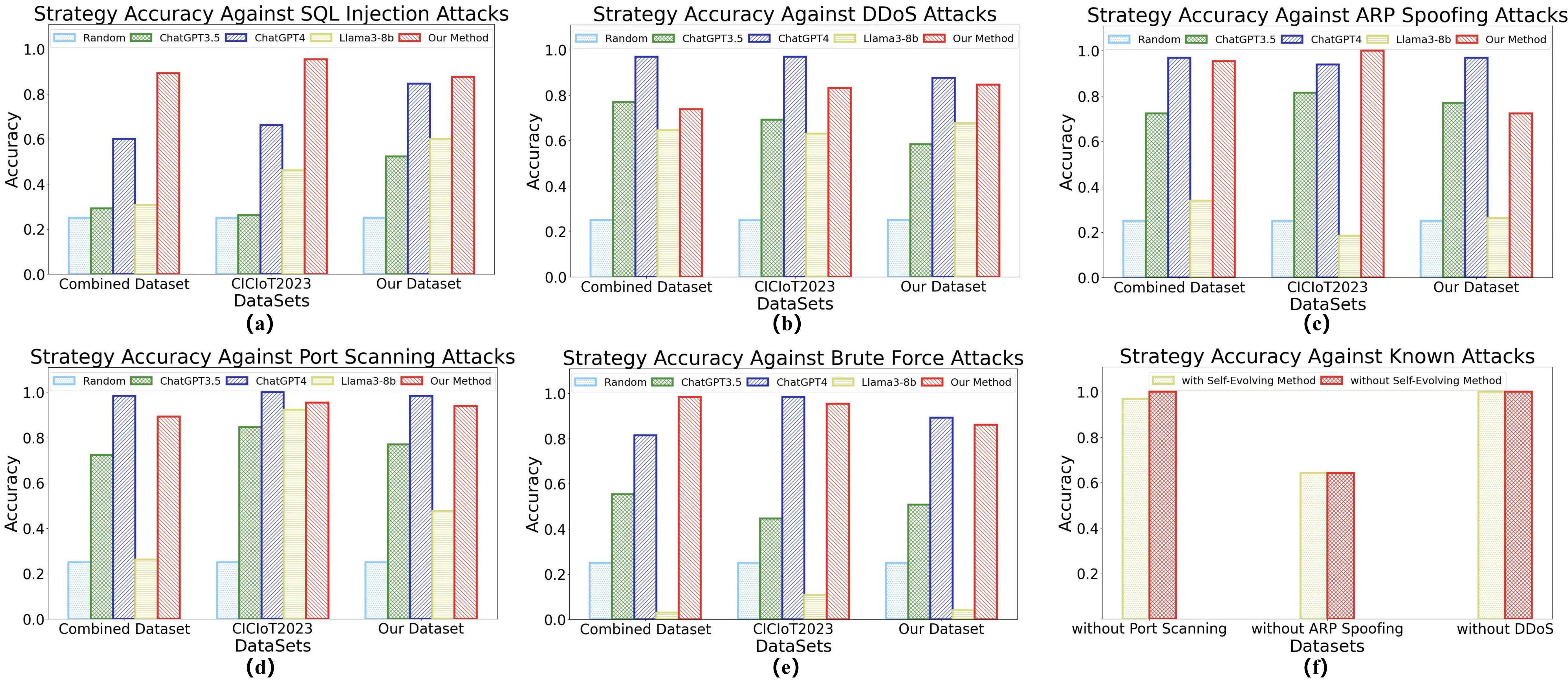}
  \caption{Illustration of Performance. (a), (b), (c), (d), and (e) compare the accuracy of security strategies generated by different LLMs against unknown threats. (f) compares the accuracy of security strategies generated with and without our 6G-INST against known attacks. Experimental results demonstrate that our security framework can generate highly accurate security strategies, and that our 6G-INST significantly improves the accuracy of security strategies against unknown threats, while maintaining the accuracy of strategies against known threats.}
  \label{fig:fig4}
  \vspace{-6mm}
\end{figure*}

\subsection{Security Strategy Generation and Parameter Update}
After real-time data collection and the prompts and threat information generation, we will generate security strategies for them using the Strategy Generation Agent. Additionally, we will periodically fine-tune the LLM with the expanded training dataset to enable its self-evolution.

\noindent \textbf{1. Strategy generation:}
Our LLM-6GNG is built using Llama3-8b because it has fewer model parameters and higher operational efficiency, enabling it to meet the real-time requirements for generating security strategies. However, In 6G-INST, we do not need to consider the efficiency of security strategy generation. Instead, we focus on the effectiveness of the generated security strategies. Therefore, we can choose a LLM with more parameters and provide additional support to assist the model in generating security strategies.

We choose GPT-4 to implement the Strategy Generation Agent in our proposed 6G-INST, and equip it with a vector database. Our 6G-INST automatically and periodically crawls security knowledge and stores it in the vector database to ensure that the content in the database contains the latest security knowledge. By utilizing the Retrieval-Augmented Generation (RAG) approach, we enhance the effectiveness of the Strategy Generation Agent, enabling it to analyze threats that are unknown to the LLM-6GNG and provide reasonable security strategies.

\noindent \textbf{2. Fine-tuning and parameter update:}
The newly generated instructions are utilized to fine-tune the Twin Specific Strategy Agent. Concurrently, we periodically update the model parameters of the Specific Strategy Agent to ensure our LLM-6GNG can handle emerging attack scenarios and threats effectively, thus enabling the self-evolution of LLM-6GNG.

\section{Experiments and Numerical Results}
\subsection{Experimental Setup}
Our network models are executed on an Ubuntu 20.04.4-powered server with 32GB of RAM, a 12th Gen Intel(R) Core(TM) i9-12900 CPU, and a NVIDIA GeForce RTX 3060 GPU. The simulation of the attack scenarios is carried out using ns-3 version 3.37. In addition, we install and optimize Llama3-8b model on a server running Ubuntu 20.04.1 with 512GB RAM, an Intel(R) Xeon(R) Gold 6248R CPU, and four NVIDIA A40 GPUs.
For instruction tuning of the LLMs, we employ the LoRA approach.

To simulate the 6G SAGINs communication environment within our 6G Simulator, we set up one GEO satellite node, 12 LEO satellite nodes, 80 UAVs, 240 ground eNodeBs, and 12,000 UEs for the sub-network. The heights for the UAV, LEO, and GEO satellites are set at 100 meters, 600 kilometers, and 35,786 kilometers, respectively.

\subsection{Datasets}
To evaluate the performance of our security framework in generating security strategies against various types of attacks across different usage scenarios, we select three datasets for experimental validation. Each dataset includes five attack types: SQL injection attack, DDoS attack, ARP spoofing attack, port scanning attack, and brute force attack. 

The first dataset is the Combined Dataset, which consists of the ARP spoofing attacks from the ARP Poisoning and Flood attack in the SDN \cite{c11} dataset, the SQL injection attacks from the CSE-CIC-IDS2018 \cite{c9} dataset, and the DDoS attacks, port scanning attacks, and brute force attacks from the CICIDS2017 \cite{c9} dataset. The second dataset is the CICIoT2023 \cite{c10} Dataset. The third dataset is collected from our 6G Simulator, which contains attacks in space-air-ground integrated networks.

\subsection{Baselines}
To evaluate the accuracy of security strategies generated by our LLM-6GNG, we compare our LLM-6GNG with Random method. The Random method serves as a performance baseline, randomly selecting security strategies. This method has an accuracy rate of 25\% in a 4-to-1 selection scenario. For the LLM-6GNG, we respectively deploy GPT-3.5, GPT-4, and Llama3-8b to verify its performance. Specifically, GPT-3.5 and GPT-4 are advanced closed-source LLMs with a large number of parameters.

To validate the effectiveness of our method, we compare the accuracy of security strategies generated by our method (Llama3-8b with our proposed 6G-INST) against those generated by GPT-3.5, GPT-4, and Llama3-8b without our 6G-INST when addressing emerging threats.
By comparing the accuracy performance of our LLM-6GNG with and without the application of our proposed 6G-INST, we aim to quantify the enhancement effect of our 6G-INST on the performance of our LLM-6GNG.

\subsection{Performance}

Figures \ref{fig:fig4} (a), (b), (c), (d), and (e) demonstrate the high accuracy of security strategies generated by our LLM-6GNG and the effectiveness of our 6G-INTS.

Firstly, we conduct a comparative experiment of the security strategy generation performance for the SQL injection attacks. As depicted in Figure \ref{fig:fig4} (a), we compare the performance of GPT-3.5, GPT-4, and Llama3-8b, which are not trained on SQL injection attacks datasets and do not employ our 6G-INST method, with our method (Llama3-8b with our 6G-INST) which is also not trained on SQL injection attacks datasets. In the experiments, we provide our LLM-6GNG with four security strategy options to choose from, one of which is the correct option.
The results show that our proposed 6G-INST significantly improves the accuracy of security strategies generated by Llama3-8b, achieving an average improvement of 45.13\% across three datasets. It is notable that we are even better than GPT-4 in terms of strategy accuracy.

Figures \ref{fig:fig4} (b), (c), (d), and (e) employ the same experimental method used in (a) to test the effectiveness of our 6G-INST against four other attack types: DDoS attack, ARP spoofing attack, port scanning attack, and brute force attack. Using the 6G-INST, the accuracy of the generated security strategies for these attacks improved by 15.39\%, 63.08\%, 37.44\%, and 89.74\% on average across three datasets, respectively. It is clearly evident from these experimental results that the application of our 6G-INST results in a substantial increase in the accuracy of the generated security strategies for all four attack types.

The five graphs discussed above show that our proposed 6G-INST can significantly improve the accuracy of security strategies against threats unknown to LLM-6GNG. The accuracy of our LLM-6GNG with the 6G-INST is notably higher than the existing Random method. Furthermore, the results demonstrate that the application of our 6G-INST achieves an average increase of 50.15\% in the accuracy of security strategies across these five attack types. 
The experiments above demonstrate that our LLM-6GNG can rely on CoT reasoning and the dynamic multi-agent collaboration mechanism to effectively analyze threat information and generate highly accurate security strategies. They also prove that our 6G-INST can effectively assist LLM-6GNG in self-evolution, enabling it to generate more accurate security strategies.

Figure \ref{fig:fig4} (f) shows that our 6G-INST does not affect accuracy against known attacks. In this experiment, we respectively exclude port scanning attacks, DDoS attacks, and ARP spoofing attacks datasets during the initial fine-tuning phase for both Llama3-8b without our 6G-INST and Llama3-8b with our 6G-INST.
Subsequently, we expose Llama3-8b with our 6G-INST to the corresponding types of attacks that were previously removed, thereby enabling the model to self-evolve against unknown threats. After the self-evolution process is complete, we test the accuracy of security strategies generated by both Llama3-8b without the 6G-INST and Llama3-8b with our 6G-INST against the known attacks. As depicted in Figure \ref{fig:fig4} (f), the accuracy rates of the security strategies generated by both models are essentially equivalent. This experiment suggests that our proposed 6G-INST does not significantly degrade the accuracy of security strategies for known attacks.

\section{Case Study}
\begin{figure*}[!htbp]
\vspace{-8mm}
  \centering
  \includegraphics[width=\textwidth]{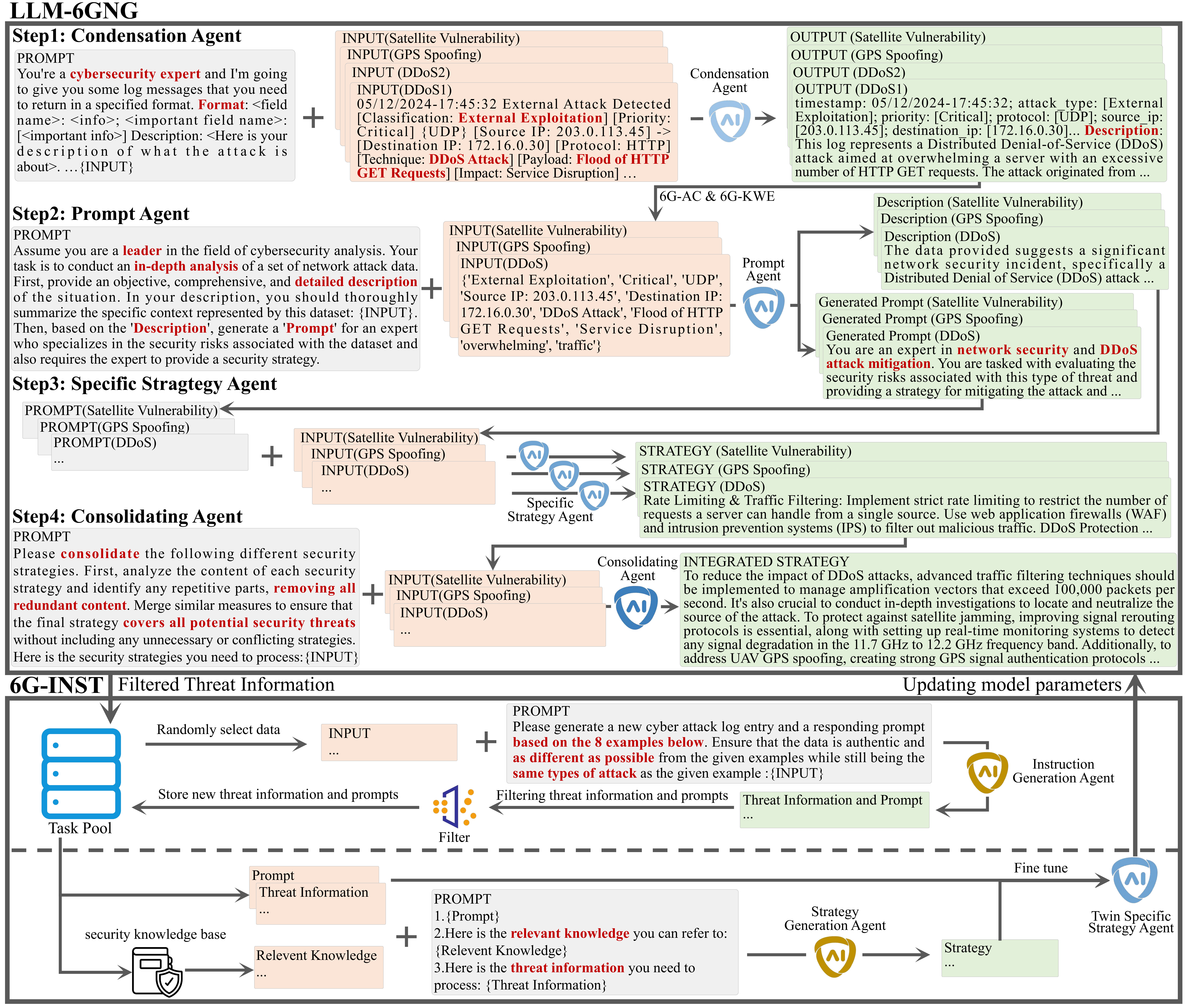}
  \caption{Case study of our security framework. This case illustrates a typical 6G SAGINs communication scenario involving threats such as DDoS attacks, UAV GPS spoofing, and satellite vulnerabilities. It details the workflows of LLM-6GNG in processing threat information and generating security strategies, and the procedure of 6G-INST assisting the self-evolution of LLM-6GNG.}
  \label{fig:fig5}
  \vspace{-6mm}
\end{figure*}

As illustrated in Figure \ref{fig:fig5}, we consider a typical 6G SAGINs scenario where a UAV transmits information to a ground station, which then relays the information to a satellite. In this scenario, we have detected threats including Satellite Vulnerability, UAV GPS Spoofing, and DDoS attacks, to demonstrate the workflow of our security framework.
Firstly, the Condensation Agent extracts key features from unstructured threat information, and generates corresponding descriptions.
Next, we perform the 6G-AC algorithm and the 6G-KWE algorithm on the output data from the Condensation Agent, thereby clustering the information and obtaining the keywords for this threat information.
Each group of key threat information is processed by a dedicated Specific Strategy Agent, which focuses solely on the key information to generate in-depth and targeted security strategies. 
Subsequently, the Consolidation Agent aggregates the security strategies produced by each Specific Strategy Agent, generates a comprehensive strategy, and returns it to the respective network.

During the operation of our LLM-6GNG, we put the actual encountered threat information that meets the requirements into the task pool of the 6G-INST, thereby expanding new threat scenarios for the fine-tuning of the Specific Strategy Agent.
In the self-evolving process, we randomly select data from the task pool and input it into the Instruction Generation Agent and Strategy Generation Agent to automatically generate new scenarios and strategies. These data are then used to fine-tune the Twin Specific Strategy Agent, and we periodically update the parameters of the running Specific Strategy Agent, completing one cycle of self-evolution.

\section{Conclusion}
This paper presents a security framework designed for multi-dimensional threat information processing and self-evolving security strategy generation in 6G SAGINs. The framework introduces two novel components: LLM-6GNG and 6G-INST, which aim to enhance the security of 6G SAGINs. Specifically, our LLM-6GNG leverages refined CoT reasoning and dynamic multi-agent mechanisms to efficiently process vast amounts of unstructured threat data and generate security strategies. 
Meanwhile, our proposed 6G-INST enables the self-evolution of our LLM-6GNG, allowing continuous adaptation to emerging scenarios and threats. 
The effectiveness of our framework has been demonstrated through experiments conducted in our 6G Simulator, and it holds promise to address the dynamic and diverse security challenges in 6G networks.

\section{Acknowledgments}
This work was supported in part by the National Key Research and Development Program of China under Grant 2022YFB2902200, in part by the National Natural Science Foundation of China under Grant 62471064, and in part by the Beijing Natural Science Foundation Program (No.L232002).

\bibliographystyle{IEEEtran}
\bibliography{ref.bib}

\end{document}